\documentclass[showpacs,aps,floatfix,10pt,superscriptaddress,pre,longbibliography,nofootinbib,preprint]{revtex4-1}
\usepackage{amsmath,amssymb,eucal,graphicx,subfigure}
\usepackage{color}
\usepackage{multirow}
\usepackage[english]{babel}

\usepackage[normalem]{ulem} 
\usepackage{setspace}

\usepackage{footnote}
\usepackage[hyperfootnotes=true]{hyperref}

\begin{document}

\definecolor{gray}{gray}{0.8}
\definecolor{darkgray}{gray}{0.6}
\newcommand{\BLUE}        [1]{{\color{blue}        {#1}}}
\newcommand{\GREEN}       [1]{{\color{green}       {#1}}}
\newcommand{\RED}         [1]{{\color{red}         {#1}}}
\newcommand{\GRAY}        [1]{{\color{gray}        {#1}}}
\newcommand{\CRIMSON}     [1]{{\color{Crimson}     {#1}}}
\newcommand{\MAGENTA}     [1]{{\color{Magenta}     {#1}}}
\newcommand{\ORANGE}      [1]{{\color{Orange}      {#1}}}
\newcommand{\YELLOWORANGE}[1]{{\color{YellowOrange}{#1}}}
\newcommand{\CERULEAN}    [1]{{\color{Cerulean}    {#1}}}

\newcommand{\BOLD}        [1]{\bf {#1}}
\newcommand{\REDBOLD}     [1]{\RED{\BOLD{#1}}}

\newcommand{\SOUT}        [1]{\RED{\sout{#1}}}
\newcommand{\FOOTNOTE}    [1]{\footnote{\RED{{#1}}}}

\newcommand{\corr}[2]{\langle{#1}|{#2}\rangle}

\newcommand{\minitab}[2][l]{\begin{tabular}{#1}#2\end{tabular}}

\newcommand{\Ca}{\ensuremath{\mbox{C}_\alpha}}
\newcommand{\Ha}{\ensuremath{\mbox{H}_\alpha}}
\newcommand{\kB}{\ensuremath{k_{\mbox{\scriptsize B}}}}


\title{Slow Normal Modes of Proteins are Accurately Reproduced across Different Platforms}

\author{Hyuntae Na}
\email{hzn17@psu.edu}
\affiliation{Computer Science, Penn State Harrisburg, Middletown, Pennsylvania, USA}
\author{Daniel ben-Avraham}
\author{Monique M. Tirion}
\email{mmtirion@clarkson.edu}
\affiliation{Physics Department, Clarkson University, Potsdam, New York
13699-5820, USA} 

\begin{abstract}
The Protein Data Bank (PDB) \cite{PDB} contains the atomic structures  of over $10^5$
biomolecules with better than 2.8\AA~resolution. The listing of the identities and coordinates of the atoms comprising each
macromolecule permits an analysis of the slow-time vibrational response of these large systems to minor
perturbations. 3D video animations of individual modes of oscillation demonstrate how regions interdigitate to create cohesive
collective motions, providing a comprehensive framework for and familiarity with the overall 3D architecture. Furthermore,
the isolation and representation of the softest, slowest deformation coordinates provide opportunities for the development of
mechanical models of enzyme function. The eigenvector decomposition, therefore,  must be accurate, reliable as well as rapid
to be generally reported upon.
We obtain the eigenmodes of a 1.2\AA~34kDa PDB entry using either exclusively heavy atoms or partly or fully reduced atomic sets;
 Cartesian or internal  coordinates; interatomic  force fields derived either from a full Cartesian potential, a reduced atomic potential or 
a  Gaussian distance-dependent potential; 
and independently developed software. These varied technologies are similar in that each maintains proper stereochemistry either by use of 
 dihedral degrees of freedom which freezes bond lengths and bond angles, or by use of a full atomic potential that includes realistic bond length
 and angle restraints.  
We find that the shapes of the slowest eigenvectors are nearly identical, not merely similar.
\end{abstract}

\maketitle
	
\section{Introduction}
\label{introduction.sec}

The softest normal modes of proteins pertain to
motilities centered on the active sites of enzymes.  
Active sites typically reside in clefts or crevasses near the surface of enzymes with the catalytic residues   juxtaposed across these clefts, with nucleophiles situated on one surface and  proton donors  situated across the crevasse on an opposing domain. The softest modes  center on these crevasses, with motilities generally pertaining to an opening/closing across the cleft; a grinding motion of one surface past the other; and a tilting motion from side to side. Remarkably, the catalytic residues inside the active clefts 
are seen to remain in fixed and constant apposition. 
These features are a consequence of the overall design and layout of the molecule,
enabled by the collective disposition of all  residues and  the defining characteristics of the softest innate degrees of freedom (dof) identified by 
  Normal Mode Analysis (NMA).

The motilities centered on  active sites but enabled by the collective atomic dispositions are thought to be robust 
 \cite{doruker03,Ming8620,sanejouand06,zheng06,echave12}. 
Freezing the dofs associated with apparent hinge residues prior to a NMA, for example,
do not alter the appearance of the low-order modes, yet the protein data bank reveals instances of structures where the re-alignment of a single residue in an active site significantly stiffens the computed lowest-order mode \cite{tirion15b}.  It seems plausible therefore that the softest eigenmodes  are relevant to function, with  motility patterns selected to fine-tune ligand entry and exit propensities.  
Identifying these intrinsic motilities therefore may help to fully characterize the structure/motility/function relationship of these systems, in addition to providing a conceptual framework for the layout and design of each macromolecule.

Despite their importance, the lowest-order normal modes are not typically reported on, nor their data recorded, with each new PDB deposition.
This might be due to a skepticism that the softest computed modes are unique and accurately identified.  Reservations about the reliability of protein NMA is perhaps also linked to an association of normal modes and
the crystallographic Debye-Waller or B factors, which purports to report on the dynamic mobility of each atom. Patterns of atomic
displacements in crystals of the same enzyme derived from independent crystallographic refinements vary \cite{artymiuk79,parthasarathy97},  and such patterns are readily reproduced with a minimum of parameters \cite{kuriyan91}.
While normal modes identify equilibrium vibrations associated with isolated  crystalline-structures, numerous other effects contribute to B factor data including  lattice vibrations, structural-ensemble averaging, crystal imperfections and model errors,
 limiting  the use of B factors  as a proxy for internal vibrations. However, each (model) structure obtains a well-defined eigenvector array, just as every structure obtains a well-defined Fourier transform, and as such the normal mode signature  provides a unique opportunity to study the dynamic consequences of the overall fold.  
As normal mode studies using coarse-grained crystal structures, such as the use of only C$_\alpha$ atoms, cannot 
 distinguish inter-atomic-interaction density changes at active sites, we here 
 examine the consistency of the atomic shapes of protein normal modes as computed by  a variety of platforms that maintain stereochemical topology of all bond lengths and bond angles.

An eigenvector decomposition requires the system to be at rest, i.e. at a local minimum of its potential energy surface. For protein molecules,
this is implemented in one of two ways: either the experimentally determined coordinates are adjusted to bring the structure
to a nearby  minimum of a standard potential energy function (classic NMA), or the potential energy function itself is defined to be a  minimum
at the experimentally determined coordinates (PDB-NMA).  The first protein NMAs adopted the former strategy; adhering to sophisticated
phenomenological potential energy formulations, the experimentally determined structures were relaxed into a nearby minimum using
some type of minimization algorithm \cite{marion}.  As minimization algorithms for large assemblies with sensitive distance dependencies are
computer intensive and inexact, and since the interest in protein NMA is often in the experimentally determined coordinates,
we suggested the use of a potential energy function designed to be a minimum at the given (PDB) configuration~\cite{ben-avraham95,tirion96}.
The so-called Hookean potentials proved to be highly effective  in describing the 
eigenspectra of systems evaluated at the same (minimized) coordinates \cite{na2015,tirion15a}.

The first PDB-NMA, on an energy-minimized configuration of G-actin in order to test for consistency with
classic NMA, did not employ Cartesian coordinates to model atomic mobility, but  instead used dihedral  angles to model 
rotations about single covalent bonds \cite{tirion93}. 
The use of these internal coordinates was deliberate: by constraining bond lengths and angles to their PDB values,  
proper stereochemical topology was maintained. In this earliest formulation, 
all (heavy) atoms moved within a potential created by the non-covalent interactions (NCI)
between atom pairs at least 3 bond lengths and no more than 5\AA~ apart. 
Initially, all NCIs  were characterized by a universal spring constant whose value was adjusted to shift the frequency spectrum to the proper register.

We subsequently studied the limits  on the precision of the PDB-NMA formulation compared
to classic NMA \cite{na2015,tirion15a}. By adjusting {\it each} NCI spring constant's magnitude in accordance to a ``parent potential'' such as CHARMM \cite{CHARMM2009} or L79 \cite{Levitt83},
and by introducing spring constants (also modeled on the same parent potential) for each bonded dihedral coordinate, the equivalence
between PDB-NMA and classic NMA is maximized: indeed, when using the same (parent) potential on the same (minimized)
coordinates, PDB-NMA and classic NMA produce nearly identical eigenspectra and concomitant eigenvectors.
(While no detailed analyses of how eigenspectra and eigenvectors shift as a consequence of minimization away from the PDB coordinates have been published to our knowledge, Na, et al. ~\cite{na2016universal} find that  the slowest modes seem unaffected by minimization while the main 50 cm$^{-1}$ peak of the eigenspectrum distribution shifts to higher frequencies.)

A different line of research  explores the consequences of {\it structural} coarse-graining combined with universal spring constants in formulating the PDB-NMA equations, 
resulting in the  field of elastic network models (ENM) that include 
the Gaussian network model (GNM) \cite{bahar97}
and 
the  anisotropic network model (ANM)   \cite{atilgan01,hinsen1998}.  
These ENM formulations are typically cast  using
a subset of atoms, such as only C$_\alpha$ atoms, and can therefore be applied to vast biomolecular assemblies such
as viral capsids, ribosomes and motor domains \cite{rader05,wang04,zheng03}. 

Here we continue to explore the question: how consistent are the various PDB-NMA formulations? Do the eigenvectors
computed using different algorithms and different parent potentials match or are they merely similar?
Na and Song~\cite{na2015,Na15c} systematically examined the limits of the accuracy of ANM and GNM versus classic NMA with different coarse-graining levels. 
Not surprisingly, the extremely simple potentials used in ENM 
lead to rapid loss of precision in comparison to classic NMA.
Instead of ENM type models,  we here examine the consistency of  PDB-NMAs that maintain stereochemical topology, either
by use of internal coordinates (dihedral angles) or by use of a full atomic, Cartesian potential energy function.

The PDB-NMA technologies here examined, sbNMA~\cite{na2014}, ATMAN~\cite{tirion15a} and ProMode~\cite{Wako04,ProMode13}, 
are similar in that each employs  the full heavy atom-set of PDB entries,
and each maintains accepted stereochemical topology, either by use of internal dihedral coordinates  that freeze bond lengths and angles 
(ATMAN, ProMode), or by use of all Cartesian dofs that properly scale each pairwise interaction (sbNMA).
They differ in significant ways. First, ProMode excludes all hydrogen atoms in the computations;
ATMAN uses hydrogen atoms only on mainchain amide nitrogens (except for proline residues and N-terminal  amides); while sbNMA 
 builds in the full complement of hydrogen atoms. Second,
the Hookean energy functional employed by sbNMA is based on
CHARMM22 \cite{Mackerell1998,Mackerell2004} and includes  all bonded and nonbonded  interactions;
the energy function used in ATMAN is derived from the united-atom L79 potential; while ProMode 
includes only nonbonded interactions modeled by a single, distance-dependent spring constant.  Third, each technology
uses slightly different criteria for inclusion of atom pairs for the NCI-set: ATMAN and sbNMA exclude NCIs between atom pairs whose distance
of separation exceeds the inflection point of their van der Waals curve, while ProMode uses a Gaussian cutoff of 5\AA.
In addition to the long-range cutoff, sbNMA excludes NCIs between atom pairs closer than three bond lengths, while ATMAN's minimum cutoff is four bond lengths 
(the interaction of atom pairs separated by four bond lengths is modeled by the dihedral stiffness constants).
And fourth,  ProMode sets the dihedral stiffness constants  k$_\phi$, k$_\psi$ and k$_\chi$ to zero and sets k$_\omega$ to an arbitrary large value~\cite{wako17};
ATMAN uses dihedral stiffness constants from the actual 2nd derivative of the L79 parent potential,
while sbNMA uses the {\it maximal} stiffness constants of CHARMM22 for all dihedral angles (as if they were at their most stable configuration).  More details are given in Section~\ref{methods.sec}, and a summary of the differences is presented in Table~\ref{platforms.table}.

These PDB-NMA technologies were derived and encoded independently and, given both the breadth of the variations 
in the formulations as well as the possibility  of inconsistencies in the complex coding algorithms,
we expected to see  differences in the appearances of the modes. And so,
to quantify the loss of accuracy introduced  by the exclusion of bond length and bond angle dofs, 
we used the TNM protocol~\cite{Mendez10} to extract from the sbNMA Hessian matrix the subset of modes belonging to  the dihedral angles.  
Here, we label this as sbNMA-TNM.
As sbNMA and sbNMA-TNM use identical energy functions and hydrogen atom subsets, differences in their modes can only be attributed to the exclusion of bond-length and bond-angle dofs.  To then account for the variable effect of the parent potential, we  compared 
sbNMA-TNM to ATMAN, since they share  the use of torsional dofs
but derive from vastly different potentials: CHARMM22 (sbNMA-TNM) and L79 (ATMAN).  
Likewise, we re-formulated ATMAN to AT-ENM in order to mimic the conditions used in ProMode: all NCI stiffness constants set
universally to one and dihedral stiffness constants set to zero. Any remaining differences between ProMode and AT-ENM 
are  due to the alternate NCI sets, and to the use of {\it all} $(\phi,\psi,\omega,\chi)$ dihedral angles in ProMode, as opposed to the $(\phi,\psi,\chi)$ only in ATMAN. 
But rather than the anticipated differences in the atomic displacements, 
the overall atomic displacement-correlations between these varied techniques for {\it the slowest modes} were very nearly perfect. The almost
perfect overlap of the slowest eigenvectors as computed by these varied techniques
points to a correct coding  as well as an unexpected
level of robustness in  the softest modes. Indeed, 
in order to pinpoint the slight atomic-displacement mismatches between these 
alternate PDB-NMA technologies, we had to rely on correlation comparisons {\it per atom} per mode. Only by examining the atomic-displacement correlations 
in this fashion were we able to observe which atom set, and to what extent, is affected by the differences in the various approaches.

This almost exact overlap of the slow eigenmodes computed by three distinct PDB-NMA protocols, sbNMA, ATMAN and ProMode, was a surprise.
While the {\it large-scale} shape of the slow modes of an object are known to be robust and largely invariant under nonrigid deformations and noise, 
we did not anticipate  these basis vectors to match so closely at an
{\it atomic} level. The exclusion of bond-angle degrees of freedom, for example, might
be expected to result in subtle but significant differences even for the slow modes. Furthermore, the inclusion of all hydrogen atoms nearly doubles
the atom-set and therefore also roughly doubles the NCI-set, thereby introducing a distinct weighting scheme in sculpting the modal shapes. 
That the computed heavy atom displacements match for the lowest frequency modes across these several platforms demonstrates
a surprising degree of robustness and increases confidence in mode shape computations for understanding
molecular dynamics and function. 
Even though different force fields have different functional forms and coarse-graining levels, they are designed to approximate a
natural phenomenon: the relaxation response in the slow time regime.
The demonstration  that low-frequency dynamics determined by different approaches are very similar, therefore,
provides further support to  efforts to include the low frequency oscillation data along with the coordinate data for PDB entries. 

In all, this work demonstrates that the effects of eigenmodes computed using different technologies may be
compared on an atom by atom basis, and hence the equivalence of alternate bases vectors  assessed.
We find that maintaining stereochemistry, either by use of 
 dihedral degrees of freedom that freeze bond lengths and bond angles, or by use of a full atomic potential that includes realistically stiff bond length
 and angle spring constants, results in reliable, reproducible slow modes.
 Furthermore, the omission of hydrogen atoms does not alter the displacement vectors of the atoms associated
with core residues, but primarily those of solvent exposed, surface residues.  The most significant visual difference between these three technologies,
sbNMA, ATMAN and ProMode, is in the rotations of side-chain $\chi$ angles; specifically how side-chains in compression zones
deform to accommodate relative re-orientation of regions.  Finally,
comparisons of the softest eigenmodes of a system computed using alternate formulations permit an explicit 
determination of the contribution of specific structural or
energetic features to the dynamic response of systems, revealing hierarchies in dynamical responses to aid in the design of rational
phenomenological potential energy formulations.

\section{Methods}
\label{methods.sec}

\subsection{PDB structure}
To examine the effect of alternate PDB-NMA protocols on computed mode displacement vectors, we sought a PDB entry reporting
high confidence in the precision of each median atomic position, as determined by a range of macromolecular refinement parameters \cite{wlodawer07}.
A glycoside hydrolase named xylanase 10A with PDBID 1E0W
is cited in \cite{wlodawer07} as such a structure,
reducing complications arising from incorrectly modeled electron density data.
1E0W consists of 302 amino acid residues, 2,333 non-hydrogen protein atoms, two disulfide bonds and  438 crystal waters~\cite{Ducros00}.
Except insofar as the development of the empiric variables used in the phenomenological potentials reflects a bias towards solvated structures,
no additional efforts were made to include solvent in the PDB-NMA analysis.

\begin{figure}[h]
\includegraphics[width=0.5\textwidth]{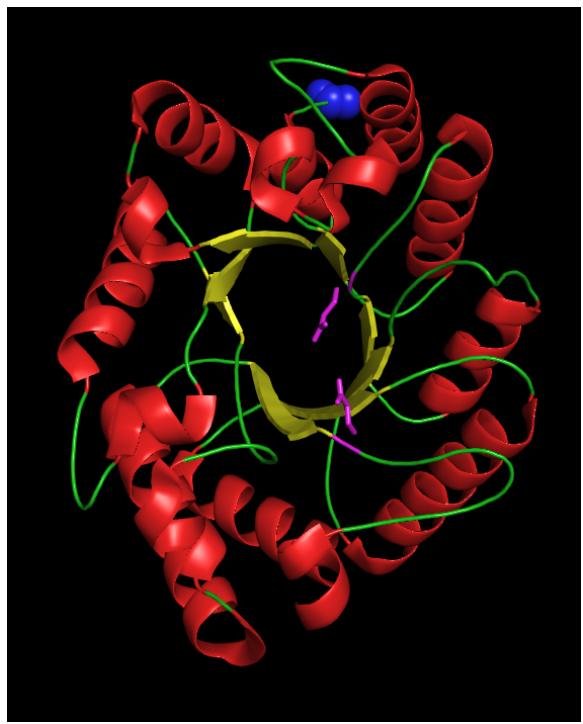}
\caption{Cartoon representation of the 302 amino acid  1E0W structure, with the eight $\beta$-strands in yellow
and the outer $\alpha$-helices in red.  The N terminal residue is  shown as blue spheres, while the catalytic
residues Glu128 and Glu236  are illustrated as magenta sticks at the bottom and top, respectively,
of the active site cleft, which runs roughly horizontal in this representation. }
\label{Figure01}
\end{figure}

As seen in Fig.~\ref{Figure01}, the single polypeptide chain  folds into a classic $(\alpha/\beta)_8$~TIM barrel fold, with the
eight parallel $\beta$-strands aligned in a tubular fashion, forming  the inner surface of the barrel. 
Eight $\alpha$-helices are arranged exterior 
and roughly parallel to the tubular $\beta$-sheet strands.  
As is typical for this fold, the active site is situated in a cleft at the C terminal end of the  barrel.
This end of the barrel  presents as the more open or flared terminus with multiple loops forming the active site cleft, while the N terminal end of the 
barrel lacks the pronounced crevasse and loop insertions and is stabilized by the presence of the two disulfide bridges.  
If the molecule is oriented as in Fig.~\ref{Figure01}  with the active site cleft
 roughly horizontal, the Glu236 nucleophile above the cleft and the Glu128 general acid/base below the cleft, then the dimensions of the molecule
are roughly 35\AA~ front to back, 50\AA~ top to bottom and 40\AA~ side to side.  

We restrict our study to this one model xylanase system, known for its excellent crystallographic resolution and whose
eigenspectra have been examined previously \cite{tirion15b}.  One reason for this is the sheer volume of information we attempt to
convey: the spread in the atomic displacement vectors of nearly 3,000 atoms due to three alternate formulations of the 
PDB-NMA formalism. In addition, experience demonstrates that the interaction-specific analyses here examined consistently
reproduce classical NMA results when performed on the energy minimized coordinates, regardless of size or fold, 
as predicated on the observed universality of the vibrational spectra of globular proteins \cite{na2016universal}

\subsection {sbNMA and TNM}

Spring-based NMA (sbNMA) is developed by simplifying the traditional normal mode analysis (NMA).
Na and Song~\cite{na2014} decomposed the NMA Hessian matrix into two terms: spring-based terms proportional to the spring constant, and force/torque-based terms proportional to the inter-atomic forces or torques.
Since the force/torque-based terms require  energy minimization while their contributions are considerably less than the spring-based terms~\cite{Na14a}, sbNMA keeps only the spring-based terms.
To ensure stability while maintaining  accuracy, sbNMA does not include electrostatic interactions and springs with negative spring constants.
sbNMA most accurately resembles the vibrational modes and frequencies of NMA without requiring a cumbersome energy minimization step.

To analyze the modal spectrum of 1E0W using sbNMA, 
2,207 hydrogen atoms were added to the 2,333 heavy atom PDB coordinates using PSFGEN \cite{VMD96} and their locations minimized using NAMD \cite{NAMD05},
resulting in a system with $3 \times 4,540=13,620$ dofs.
To set up the all-atom sbNMA Hessian matrix,
the CHARMM22~\cite{Mackerell1998,Mackerell2004} force field was used to parameterize the energy profiles of  12,746 proper dihedrals, 836 improper
dihedrals, 4,591 bond lengths, 8,198 bond angles and 26,007 noncovalent interactions
as described in Na and Song~\cite{na2014}.  The eigendecomposition was accomplished using MATLAB.

The full-atomic, full-Cartesian Hessian matrix developed by sbNMA may be used to study the isolated contributions of
particular internal dofs.  
To isolate from the  sbNMA Hessian matrix the contribution of only the dihedral dofs while constraining the
bond-lengths and angles, we used
the TNM matrix projection algorithm of Mendez and Bastolla ~\cite{Mendez10} (code developed in-house by Na).
The resultant reduced  sbNMA-TNM analysis includes  $1,684$ proper dihedral dofs for the rotations about  single bonds.  
The use of this subset of the sbNMA Hessian matrix guarantees that differences in the computed sbNMA-TNM eigenvectors compared to those 
obtained from the full sbNMA Hessian matrix are solely due to the exclusion of bond length, bond angle and improper torsional angle dofs. 
This permits a direct evaluation of the relative contribution of the softer and stiffer coordinates to the low-order modes.

\subsection{ATMAN}

For the ATMAN analysis,
the  program REDUCE \cite{Word99} is used to build hydrogen atoms on main-chain amide hydrogens.
The resultant 2,627 atom-system obtains $2\times302$ $\phi$ and $\psi$ angles, less the 7 proline residues and $\phi_1$, 
or 597 main-chain dofs, as well as 
505 $\chi$ angles for side-chain vibrations about single bonds, for a total of 1,101 dofs. 
The energy of each pair of $(\phi,\psi)$ angles of every residue is calculated as the sum of 4 Gaussian-like potentials:
$\Sigma_{i=1}^4 E_i g(\phi-\phi_0^i,w^i_\phi)g(\psi-\psi_0^i,w^i_\psi)$,
where $g(x,w) = \exp [-0.693(1- \cos(x))/ \sin(w/2)]$.  The parameter values for $E_i$, $(\phi_0,\psi_0)$, $w_\phi$ and $w_\psi$
are given in Table 4 of Ref.~\citenum{Levitt83}. Side chain $\chi$ dihedral energies are calculated using $K[1+\cos(n\chi+\delta)]$ with 
the $K$, $n$ and $\delta$ parameters, dependent on the identity of the central bonded pair of atoms,  given in Ref.~\citenum{Levitt83}.
In addition,
there are  12,215 NCI between atom pairs more than 4 bond lengths apart and less than the distance of the inflection
point of their Lennard-Jones curve, again using the parameterization of Ref.~\citenum{Levitt83}.   
(Beyond the inflection point, the spring constant computed from the parent potential by the ATMAN and sbNMA algorithms become negative.) 
The average NCI per atom is 9.30.
The Hessian and mass matrices are computed analytically and coded in Fortran79, and the generalized eigenvalue equation is diagonalized
using LAPACK \cite{LAPACK99}.  

To test for consistency between ATMAN and ProMode,
we altered the parameterization of ATMAN to better match the conditions used in ProMode elastic: we set all dihedral stiffness constants close to zero (0.5 kcal mol$^{-1}$), set all noncovalent interaction stiffness constants to 1 kcal mol$^{-1}$, 
and included all NCI between atom pairs more than 3 bond lengths apart, rather than 4.  
This parameterization is labeled AT-ENM.  
While this parameterization creates a closer match to ProMode (next section) than  ATMAN, differences remain,
including the noncovalent interaction set used, as well as 
 the distance-dependence used  to scale the
strengths of the NCI bond strength constants. However, it is expected that a closer correlation exists between
ProMode and AT-ENM  than ATMAN.

\subsection{ProMode}

Since 2003, protein data bank Japan (PDBj.org), a  worldwide protein data bank (wwPDB) partner, provides
ProMode elastic modal analyses of a large and growing number of PDB entries~\cite{Wako04,ProMode13,wako17}.
Initially configured to minimize coordinates to a nearby minimum of the potential energy surface, the current
implementation of ProMode employs a GNM-based PDB-NMA~\cite{wako17} with the NCI
spring constants between atoms $i$ and $j$ scaled according to $k\exp[ -(r^{0}_{ij}/a)^2 ]$ where
$r^0_{ij}$ is the PDB distance between atoms $i$ and $j$,  $k = 1.0\,{\rm kcal}\,{\rm \AA}^{-2}{\rm mol}$ and $a = 5.0\,
$\AA~ for all atom pairs. (The Gaussian decay of the spring constant means that interactions are negligible beyond a distance of order $a$ --- this is mimicked in AT-ENM with a constant spring value and a sharp cutoff.)
ProMode excludes all hydrogen atoms in the analysis, even when present in the PDB entry, and
uses the dihedral coordinates of main-chain $\phi$,  $\psi$ and $\chi$ angles, with a zero spring constant, as
well as the planar peptide $\omega$ dihedrals, but with a stiff spring constant of $K_\omega=10\,{\rm kcal\,mol^{-1}\,rad^{-2}}$~\cite{wako17}. In addition to animation sequences, plots of time averages and correlations of fluctuations of
atoms and dihedral angles, the eigenvector entries per atom per mode for the 10 lowest-order modes are
available for download at \url{http://pdbj.org/promode-elastic/1e0w}.

\subsection{Comparison  of Eigenmodes}

Hessian matrices developed using different dofs, such as Cartesian coordinates versus dihedral rotations, cannot be directly compared, 
and hence it is
necessary to compare the effects that the computed eigenmodes have on the PDB coordinates.
To facilitate comparisons,
 it is  convenient to express the modes in a uniform format.  This is achieved by activating each mode to a fixed amplitude (corresponding to either a particular thermal activation temperature, or to a preset RMS displacement) and recording the changes $\Delta{\bf r}_a$ experienced by each individual atom $a$ from its PDB original position, in Cartesian coordinates.  Thus, each mode ${\bf U}_i$ is now a vector consisting of the $\Delta{\bf r}_a$ displacements. The ProMode data entries with file extensions flcatmE in PDBj.org provide a useful format model for this purpose.  
Once scaled to the same magnitudes, or RMS displacements, simultaneous animations of particular modes
permit a {\it direct visual comparison} of the shapes of the modes as computed by the different technologies.

Limits on visual acuity, however, limit the precision with which differences may be discerned, and
so we turn to  more quantitative measures to quantify subtle differences.  First, we compare the {\it RMS displacements} of each C$_\alpha$ for equivalent modes of different technologies.  This provides an overall measure of mobility (e.g., atom $i$ is twice as  mobile in mode 1 of technique~A, than in technique~B), but ignores the direction of motion of each atom.  The RMS comparisons are meaningful because of their immediate relation to the experimentally available temperature factors.  Second, we compute the {\it overall correlations} between mode~$i$ of technique~A (${\bf U}_i$) and the analogous mode~$i$ from technique~B (${\bf V}_i$):
\[  
\langle{\bf U}_i|{\bf V}_i\rangle=\frac{{\bf U}_i\cdot{\bf V}_i}{\|{\bf U}_i\|\,\|{\bf V}_i\|}\,.
\]
A correlation of 1 indicates perfect agreement between the two modes.  We also check the cross-correlations between {\it different} modes ($i$ and $j=i\pm1$) from different techniques, $\langle{\bf U}_i|{\bf V}_j\rangle$. Orthogonality of the modes implies that $\langle{\bf U}_i|{\bf U}_j\rangle=0$ within each technique, and the goal is to see to what extent orthogonality carries over between modes from {\it different} techniques.

Overall correlations between analogous modes may be very close to 1, yet such high correlations may hide some large differences for a small (but significant) subset of the atoms.  Suppose, for example, that a $10,000$-atom protein exhibits 0.98 correlation between mode 1 of techniques A and B ($\langle{\bf U}_1|{\bf V}_1\rangle=0.98$).  This could be achieved with 99\% of the atoms moving in perfect correlation, and a small minority of 100 atoms moving exactly out of phase (in opposite direction to one another, in the two modes).  To test the actual origin of the discrepancies we look at the individual{ \it atomic correlations} between the displacement of atom $a$, $\Delta{\bf r}_a$, in the mode from technique A versus the same displacement in the mode from technique~B.  Indeed, this detailed measure reveals that the discrepancies in the overall correlations stem from disagreements among small subsets of atoms, rather than being equally distributed across all atoms. This level of detail  helps to identify the actual causes for the differences in the modes from different technologies.

\section{RESULTS}

Table~\ref{Figure012} presents the eigenfrequency data and correlations between modes computed with the techniques described in the Method section: sbNMA, sbNMA-TNM, ATMAN, AT-ENM and ProMode.
The diagonal blocks list the mode frequencies, in inverse centimeters, of modes $i$ for $i=1,\dots,5$ for each technique.
sbNMA and ATMAN  obtain nearly identical frequencies for the first two eigenmodes, even though these values are derived using vastly
differing technologies. Interestingly, the mode frequencies of sbNMA-TNM have uniformly increased by 10\% for all 5 modes, relative to sbNMA, 
indicating a stiffening of the generalized spring constants (due to the constrained bond lengths and bond angles). This trend is not seen when comparing ATMAN and AT-ENM, where the mode
frequencies shift up and down by a few percent. The mode frequencies
reported here are in line with analyses based on standard (energy-minimized) NMA for proteins of this size~\cite{na2016universal}.
On the other hand, the energy function shaping
the eigenmodes in ProMode is based on an ENM-type potential that does not maintain realistic eigenfrequency
magnitudes: the amplitudes of the modal oscillations are specified directly and arbitrarily, rather than by use of the
thermal energy factor $\frac{1}{2}\kB T$, and  ``frequencies'' are derived from these amplitudes. 

The block entries above the diagonal  of Table~\ref{Figure012} list the all-atom correlations $\corr{{\bf U}_i}{{\bf V}_i}$ of eigenvectors $i$ between the
various techniques, for $i=1,\dots,5$.  Mode 1 obtains the highest correlations across the various techniques, demonstrating
that  the lowest-order mode is most robust.
The nearly perfect correlation between sbNMA and sbNMA-TNM of 1.00, 1.00, 1.00, 0.99 and 0.98 demonstrates that the slowest, softest modes are modulated
very nearly exclusively by fluctuations of bond rotations, and not by bond angle nor bond length fluctuations.
The weakest correlations for the slowest modes are between ATMAN and ProMode, with correlations of 0.92,
0.87, 0.81, 0.71 and 0.59. 
While still strongly correlated, we
 wished to demonstrate that adopting a parameterization of ATMAN similar to ProMode improves these correlations. And indeed, 
when the parameterization of ATMAN is altered to AT-ENM (see the Methods section),  the correlations 
 increase to 0.97, 0.93, 0.90, 0.79 and 0.66.
Except for sbNMA and sbNMA-TNM, the correlations involving mode 5 are low, indicating that the higher-order modes are
sensitive to the specific parameterizations.

The blocks below the diagonal of Table~\ref{Figure012} list the first off-diagonal entries of the cross-correlations $\corr{{\bf U}_i}{{\bf V}_{i\pm1}}$, with
both off-diagonal entries reported.  Magnitudes of 0.0 signify strict orthogonality of the two vectors, with no
overlap, while matching off-diagonal entries demonstrate symmetry in the deformations being compared.
Deviations from 0.0 as well as asymmetries in the off-diagonal pairs provide independent assessments in reliability.
While these all-atom averages preclude detailed interpretations, there is broad agreement with the trends seen above the diagonal.
The low-order eigenvectors computed via sbNMA and ATMAN are largely orthogonal, and also here the agreement seems
stronger between sbNMA and ATMAN, rather than sbNMA-TNM and ATMAN.  The correlations with ProMode
improve, as before, with the ENM-type parameterization of AT-ENM. In all, these data demonstrate broad agreement
with no obvious errors or inconsistencies between these varied techniques.  

Fig.~\ref{Figure05} shows the root mean square deviations, RMSD, in \AA, at a temperature of 20$^\circ$ C, per $\Ca$ atom for modes 1 (top), 2 (middle) and 3 (bottom), 
derived from sbNMA (black), ATMAN (blue) and ProMode (dashed orange). The location of each of the eight $\beta$-strands (B1-B8) is marked,
and the two catalytic residues Glu128 and Glu236 are indicated by magenta markers.
In theory, the scale of each curve depends solely on the eigenfrequency, however as the three data sets were derived using independent orthonormality and therefore scaling formulations, it was easiest to align the results of sbNMA and ProMode to those of ATMAN, by matching each curve's average value to those of ATMAN.

The relative RMSD predictions due to mode 1 match closely, with every peak and valley reproduced by each technique and the width of each peak 
likewise well reproduced.
The predictions of ProMode differ slightly from those of sbNMA and ATMAN in the height of peaks, with several regions either obtaining higher estimates, like
the peaks in the residue range 100-200, or somewhat lower estimates, as the peaks at 55 and around 270-280. Mode 2 RMSD values match slightly less well,
with the surface exposed region connecting B2 to B3 obtaining a broader peak with ATMAN and a narrower peak with ProMode in comparison to sbNMA, and with
the active-site B3 to H4 loop obtaining a narrower peak with ATMAN in comparison to sbNMA. Furthermore, both sbNMA and ProMode predict a pinning of
solvent exposed residues Asp132 and Ser134 in the lower ``lip'' of the active site cleft that is not reproduced by ATMAN. ProMode predicts slightly greater
mobility of the C and N terminal residues as well as a greater motility of the regions adjoining the cysteine bridges around residues 191--196 and 
residues 245--252. Mode 3 RMSD values obtain lower correlations and present with  a more fragmented appearance as the correlation length
of the motion decreases with mode number.  ATMAN predicts higher mobility at the N and C terminal regions and lower mobility for the remainder of
the chain relative to ProMode and sbNMA.  ProMode correlates well with sbNMA, with mismatches in the heights of peaks, not their locations.

The close  RMSD values predicted by the different technologies for modes 1--3 is striking; an observation 
supported by the visual inspection of the animation sequences of each of these modes. The animations demonstrate that
mode 1 pertains to a ``chewing'' type motion across the active site cleft;
mode 2 pertains to a ``grinding'' type motility along the active site cleft; while mode 3 pertains to a side-wise tilting along the active site cleft.
Differences in the appearance of the modes by different techniques cannot be detected in main chain traces of the simultaneous animation-sequences of modes 1,
2 or 3.  Careful examination of the  full-atomic 3D-animations demonstrates slight differences in how the side chains near hinge regions
``yield'' to accommodate relative motions.
Fig.~\ref{Figure06} plots the histogram of the $\chi$ angle distribution of mode 1, relative to the PDB orientation, for the 505 single-bond
(non-resonant) dihedral $\chi$ angles in 1E0W. While the vast majority of the side-chain dihedrals rotate by less than 1$^\circ$ for mode 1,
at higher angles the three technologies, sbNMA, ATMAN and ProMode,
 demonstrate distinct behavior, with sbNMA permitting the least $\chi$ mobility
and ATMAN the most.  The maximum deformation in mode 1 for any $\chi$ angle, for example, is less than 4$^\circ$ for sbNMA,  6$^\circ$ for ProMode and nearly 15$^\circ$ for ATMAN. The side chains experiencing the largest deformation in mode 1 according to the ATMAN
analysis are the $\chi_2$  of Asn170 and Asn173,  with the amide side-chains of the buried 
Asn170 and the surface Asn173 at the 
end of B5  reorienting to accommodate the active cleft oscillations, predictions not made by sbNMA and ProMode that obtain less than
1$^\circ$ and 3$^\circ$ respectively for these $\chi$ angles.  Interestingly, these asparagines straddle the active site
residues, Glu128 and Glu238, as they accommodate the relative reorientation of the cleft domains.

To further examine the relative dynamic consequences of sbNMA, ATMAN and ProMode,
we plot the atomic correlations (the cosine of the angle between  the atomic vector displacements) per atom and per mode.  
This measure complements the atomic RMSD values as it reports on the relative {\it orientation}
of the atomic displacements. These measures are also far more sensitive than the averaged values presented in
Table~\ref{Figure012},
and therefore  useful to correlate causes (energy gradients or forces) and effects (dynamic signatures).
Fig.~\ref{Figure09} presents the  correlations of the 2,333   atoms of 1E0W for mode 1 (top), mode 2 (middle) and mode 3 (bottom) 
between sbNMA versus ATMAN (black), sbNMA versus ProMode (red) and ProMode versus ATMAN (blue). The atom index
at the bottom of each figure runs over all heavy atoms in the PDB entry 1E0WA, while the residue index at the top of each
figure has the eight $\beta$-strand locations indicated. For clarity, the black sbNMA-ATMAN data are plotted in-phase (near correlation 1) while the comparisons between sbNMA-ProMode and ProMode-Atman are shown out-of-phase (near correlation -1).

Several features stand out in Fig.~\ref{Figure09}.  The high  all-atom (averaged)  correlations as given in Table~\ref{platforms.table} can hide differences.
The overall correlation of 0.97 between sbNMA and ATMAN for mode 1, for example, corresponding to an average alignment 
of displacement vectors of  14$^\circ$, is due to a distribution  that includes numerous exact matches  as well as a number
of less well-aligned displacement vectors. The two lowest correlations for this data set are at atom indices 244 and 1670,
corresponding to Arg36 CB on the solvent-exposed surface of the N terminal H2 $\alpha$-helix,
and Tyr214 CE2,  just prior to H7, with its phenol CD2 and CE1 atoms again solvent-exposed. These two atoms obtain 
 displacement vectors that are  nearly orthogonal according to sbNMA and ATMAN.
Further features evident in Fig.~\ref{Figure09} include the observation that those
atoms obtaining lower correlations tend to group, with similar groupings  for all three comparison sets {\it per mode}.
The groups of atoms that obtain lower correlations differ from mode to mode; so for example both atoms with the lowest
correlations in mode 1 obtain correlations close to 1 in modes 2 and 3.  Examining the locations of these groups in PyMol, we
see that the groups tend to pertain either to surface features or to  atoms situated in hinge regions where domains compress.

It is apparent that the dynamic consequences of particular energy parameterizations may be carefully scrutinized with PDB-NMA.
One can identify atom sets with the lowest correlations per mode, for example,
discard pairs that obtain vanishingly small displacements, and so filter the correlation arrays for the lowest matches. For the
data set pertaining to the atomic correlations of mode 1 entries between sbNMA and ATMAN, for example, elimination of atoms
with a  correlation higher than 0.97, or 14$^\circ$, and further elimination of  any remaining entries where both atoms' displacements
are less than 20\% of their maximum displacements, reduces the 2,333 atom set to  122  with the lowest fits.  Excluded by
this criterion, for example, are the atomic indices 244 and 1,670 belonging to Arg36 CB and Tyr214 CE2, the atoms mentioned
before as having the lowest correlations, as both atomic entries obtain  vanishingly small displacements in mode 1.
The subsets of atoms with poorer alignments may be examined to better understand what features in the various
formulations seem to drive the distinct dynamic responses. 
Fig.~\ref{Figure10} shows a surface representation of 1E0W  with catalytic residues Glu128 and Glu136 colored magenta and the 122
atoms with high motilities and lower correlations between sbNMA and ATMAN in red (not all 122 entries are visible in this
representation, but the preponderance of surface residues in this subset is demonstrated).
The 122  atoms 
belong to 52 residues; 21 of these residues have long, charged side-chains with three or more $\chi$ angles, 17 have neutral polar groups,
and only cysteines and histidines not included. In addition, 
49 of the 122 entries pertain to main-chain atoms, 18 pertain to CB atoms and 55 pertain to side chain atoms. 
Further efforts to categorize the dynamic consequences of particular energy parameterizations is ongoing and will aid in the development
of rational, minimal energy parameterization sets for fast and efficient dynamic modeling.

\section{DISCUSSION}

We  addressed the question: do the subset  of PDB-NMAs that maintain stereochemical topology 
compute identical eigenvectors in the slow-time regime, independent of  significant differences in their formulations?
As comparisons of alternate NMA formulations directly from the Hessian matrix is challenging,  we opted to compare the {\it effects}
of eigenvectors on the PDB coordinates. 
In particular, we examined to what extent normal modes analyses that reduce the size of the eigenvalue decomposition
by use of a subset of all internal degrees of freedom, namely the dihedral  coordinates, match analyses using all internal dofs.
We also examined to what extent the exclusion of hydrogen atoms, not typically reported in PDB entries, affect outcomes.
We finally examined to what extent different pairwise potential energy expressions affect outcomes. 

The development of intrinsic dihedral potentials for protein structure analysis  \cite{scheraga66,scheraga84}
grew out of the observation that bond lengths and angles are much stiffer than
bond rotation dofs, and
permitted the earliest protein NMAs to be performed using this subset of internal dofs \cite{Go83,LSS,brooks83}.
It was anticipated that the slow modes computed using all internal dofs would be very similar.
Bond length ($b$) energies are typically parameterized by the expression $K_b(b-b_0)^2$ and bond angle ($\theta$) energies
by $K_\theta(\theta-\theta_0)^2$. 
L79, the parent potential of ATMAN,  sets $K_b$ to either 250 or 500 $\mbox{kcal}\mbox{ mol}^{-1}$ and  $K_\theta$ to either 60 or 120 $\mbox{kcal}\mbox{ mol}^{-1}\mbox{rad}^{-2}$, while the average dihedral stiffness constants for $\phi, \psi, \chi$ of 1E0W are 1.75, 1.12 and 8.56 $\mbox{kcal}\mbox{ mol}^{-1}\mbox{ rad}^{-2}$.
%
CHARMM22, the parent potential of sbNMA, sets
$K_b$ between 198 and 650 $\mbox{kcal}\mbox{ mol}^{-1}\mbox{ \AA}^{-2}$ (average of 349) for 1E0W, and 
$K_\theta$ between 15 and 160 $\mbox{kcal}\mbox{ mol}^{-1}\mbox{ rad}^{-2}$ (average of 47), with
the dihedral stiffness constants between 0.2 and 28.1 $\mbox{kcal}\mbox{ mol}^{-1}$ (average of 5.1) including all $\phi$, $\psi$, and $\chi$.
In other words, bond length dofs are roughly two orders of magnitude stiffer and bond angle dofs are about one order of magnitude
stiffer than bond rotation dofs.  Is this sufficient to justify their exclusion in analyzing the slow time regime of protein mobility?
Our work comparing  the results of sbNMA and sbNMA-TNM is unambiguous: the first five
eigenvectors remain very nearly perfectly orthonormal in Cartesian space and differences in the eigenmode shapes cannot be discerned in animations. 
On the other hand, the {\it eigenvalues} of sbNMA-TNM are uniformly  stiffer, by about 10\%, compared to sbNMA. This is not surprising, since TNM
enforces infinite stiffness on the bond length and angle coordinates, leading to a stiffer response.

It was expected that the results of ATMAN would more closely parallel those of sbNMA-TNM rather than those of sbNMA, since
both ATMAN and sbNMA-TNM use the dihedral parameterization of the coordinates.  However,  the computed eigenfrequencies 
of ATMAN more closely match those of sbNMA, in fact nearly exactly, for the first two modes. Furthermore,
neither the  $\corr{{\bf U}_i}{{\bf V}_i}$  nor the
$\corr{{\bf U}_i}{{\bf V}_j}$ correlations between ATMAN and sbNMA or sbNMA-TNM point to any significant improvement in the comparison
of ATMAN to sbNMA-TNM.  That the frequencies of
the slowest modes match exactly between ATMAN and sbNMA is interesting, and suggest a frequency compensation between  
the freezing of bond lengths and angles and the exclusion of hydrogens  in L79.
 According to the CHARMM energy parameterization, bond angles involving hydrogens are softer (have smaller spring constants) than those involving only heavier atoms. As hydrogens represent nearly half the coordinates
for a fully reduced protein, freezing these softer degrees of freedom appears to translate to a stiffer harmonic response. A PDB-NMA analysis
using CHARMM19, the last parameterization permitting reduced coordinates, could confirm this hypothesis.

While the frequencies of ProMode cannot be compared to those of sbNMA or ATMAN as ProMode motilities are derived using nonphysical
energy parameterizations, the resultant eigenmodes  compare favorably, especially between ProMode and sbNMA,
 with correlations of
0.93, 0.92 and 0.90 for the first three modes, and with somewhat lower correlations between ProMode and ATMAN, of 0.92, 0.87 and 0.81.
As ProMode models noncovalent interactions with a universal (Gaussian) spring constant, this demonstrates the importance of the
use of the physically relevant dihedral coordinates to model soft deformations. 
In order to examine the source
of the lower correlation between ProMode and ATMAN compared to sbNMA, we adjusted the ATMAN parameterization
to more closely match ProMode's:
we set all NCI stiffness constants to 1.0 and increased the NCI atom-pair set by including noncovalent interactions between atoms 3 bond lengths apart, rather than
the 4 used by ATMAN. While this increases the number of NCI to 18,999 from 12,215 and the average number of NCI per atom
to 12.5 from 9.3, solvent exposed surface sidechains remained insufficiently bound to permit the exclusion of dihedral stiffness constants:
setting $K_{dihedral}$ to zero results in multiple slow modes obtaining vanishingly small RMS values as various unbound side-chains
``spin'' freely. We therefore set $K_{dihedral}=0.5$ kcal mol$^{-1}$ rad$^{-2}$.
(Increasing the cutoff distance to increase the number of NCI pairs was not attempted, as
this results in rapid loss of the universal eigenspectrum signature for proteins~\cite{tirion15a,na2016universal}.)
These fairly simple modifications  improved the correlations for the first three modes  between ProMode and ATMAN to 0.97, 0.93 and 0.90 
 --- as good or better than the correlations between sbNMA and ATMAN. 
Remaining differences between ProMode and AT-ENM are most probably due to  remaining different selection criteria for the
NCI set; the Gaussian decay of the spring constant for NCIs in ProMode vs.~the step-function cutoff in AT-ENM; as well as  inclusion in ProMode of planar $\omega$ dihedral angles (albeit with a stiff spring constant) as opposed to AT-ENM that freezes those angles.

While the slowest eigenmodes describe identical deformations of the mainchain polypeptide trace, close examination reveals 
certain subsets of sidechains behave rather differently. The vast majority of sidechains
move {\it en masse}, rigidly (with $\Delta\chi$ nearly equal to 0) as part of  larger domains. However,  each (slow)
eigenmode parses the structure into different domains,  or subsets of residue-groupings, that move {\it en masse}. 
The topological features that permit the different groupings or domains to achieve relative, inter-domain motility include
the responsiveness of that set of $\chi$s situated in  compression zones and hinge regions.  In this regard, the predictions
of ATMAN and sbNMA  differ, with ATMAN ascribing greater $\Delta\chi$ values to this subset of $\chi$s.  
The identification of residues having $\Delta\chi$ values larger than some cutoff in ATMAN, such as 5$^\circ$,  serves as an independent
tool to identify compression zones and hinge regions, and may thereby help to group residues into domains that move
as a block.  It must be stressed, however, that while the greater sidechain flexibility seen in the  ATMAN  analyses aids in discerning domains and their motilities,  experimental verification of the motility of individual, internal sidechains is lacking, though {\it in silico} MD experiments might
support one view or another.  

While comparisons of eigenmodes demonstrate that, across platforms, the slowest modes are robust at atomic resolution,
significant differences in mode shapes occur after the first five modes in the different formulations here examined.
Conclusions  derived from the higher-frequency modes, including those belonging to the main peak around $50\,{\rm cm}^{-1}$, 
 need further examination: how do the sparser number of modes of
a dihedral analysis like sbNMA-TNM intersperse within the eigenspectrum of a full-atom analysis like sbNMA? 
How does the inclusion or exclusion of some or all hydrogen atoms affect the eigenspectrum for higher-frequency modes?
We continue to study these matters.

\section{CONCLUSION}

sbNMA, using the full complement of internal, Cartesian coordinates and the sophisticated CHARMM22 force field that includes the contributions of all hydrogen atoms, likely provides the most accurate PDB-NMA results and therefore establishes the highest standard for comparison, not only for the low order modes, but for vibrations across the entire eigenspectrum.   Different levels of coarse-graining retain the precision of sbNMA to varying extents.   
The reduction of the full complement of internal dofs to exclusively dihedral rotations, while retaining  identical force field expressions for the remaining bonded as well as nonbonded pairwise interactions via sbNMA-TNM, demonstrates that the bond length
and bond angle dofs essentially do not contribute to the character of the lowest order modes. Analyses directed toward characterizing global motilities, in other words, do not benefit from the inclusion of all internal dofs. 
ATMAN, developed according to an energetically realistic dihedral potential but with use of reduced coordinates,
matches sbNMA accurately for only the first three modes. The  {\it global} dynamical character of large biomolecules, in other words, is surprisingly
insensitive to the inclusion of hydrogen atoms. ProMode, retaining the use of dihedral dofs while simplifying the pairwise nonbonded interactions through  use of a universal, Gaussian stiffness constant, and eliminating  all hydrogens,  still obtains a lowest order
correlation of 0.93 against sbNMA.  Loss of precision, in other words, proceeds inversely with frequency, with more sophisticated
formulations retaining realistic higher frequency modes while simplifications sacrifice the accuracy of higher order modes.

The near-exact overlap of the computed atomic displacements of the thousands of atoms under tens of thousands of constraints (in the form of pair-wise spring constants)
for the lowest order modes, demonstrates that each technology, ProMode, ATMAN and sbNMA, is formulated correctly,
free of errors.
However, each technology is accurate to varying precisions due to the limits on the accuracy of the
force fields employed: the cruder potentials describing only the slower or slowest mode(s) accurately. 
On the other hand, this technology also demonstrates a means to determine a minimal number of force field parameters  necessary to achieve
a given objective: if interest is confined to the slowest, longest-correlation length motions, for example,
L79 with fewer than 60 adjustable constants in its (dihedral) force field parameterization
provides the same description of the slowest modes as CHARMM22 which uses many more adjustable parameters.
Insofar as a reduction in the parameterization simplifies the isolation of those features giving rise to particular patterns, the use of these cruder potentials may be useful.

In sum, we find that normal mode analyses that maintain stereochemical topology of the thousands of atoms comprising  biomolecular systems identify {\it unique} deformation coordinates for the softest vibrations. For the TIM-barrel structure of xylanase 10A these include a Pacman-type opening/closing of the binding domain; a grinding motion across the lower and upper surfaces of the binding domain;  and a side-wise tilting motion of the binding site crevasse.  The consistent and reproducible characterization of the global motilities  that center on  active sites (in the case of enzymes), permit elucidation of those features that contribute to these motilities, and the possible alteration of these features to achieve specifically altered motility patterns.

The demonstration that vastly divergent formulations of PDB-NMA results  in nearly identical atomic displacements for the first two or three modes increases confidence in the internal consistency of these formulations to correctly identify the slow modes of biomolecular systems. Given their usefulness both to demonstrate the ``sense'' of particular protein folds, as well as their ability to motivate mechanical models suitable for experimental studies, the inclusion of low order PDB-NMA eigenvector data 
with PDB depositions significantly augments user utility.

\clearpage

\clearpage

\begin{table}[h]
\includegraphics[width=0.9\textwidth]{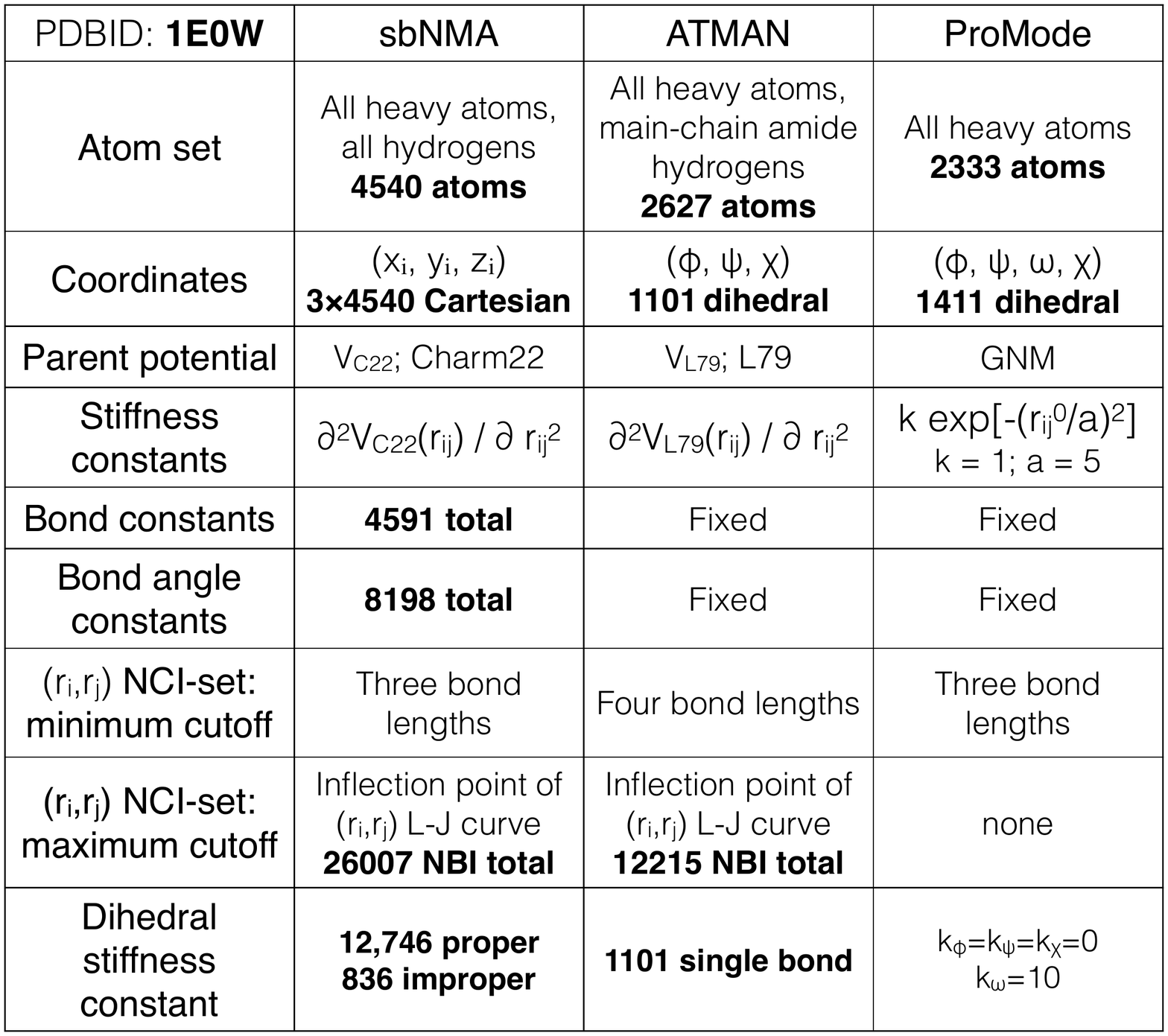}
\caption{Comparison between three different platforms used to compute PDB-NMAs that maintain standard stereochemical topology: sbNMA, ATMAN and ProMode.  Parent potential refers to the source for each atomic pairwise stiffness constant, with GNM referring to the Gaussian elastic model. Only sbNMA includes bond length and bond angle dofs. The noncovalent interaction (NCI) list of nonbonded interacting atom pairs has both a minimum and maximum cutoff criteria, with L-J referring to the Lennard-Jones curve for each pairwise  interaction.}
\label{platforms.table}
\end{table}
\clearpage

\begin{table}[h]
\includegraphics[width=0.9\textwidth]{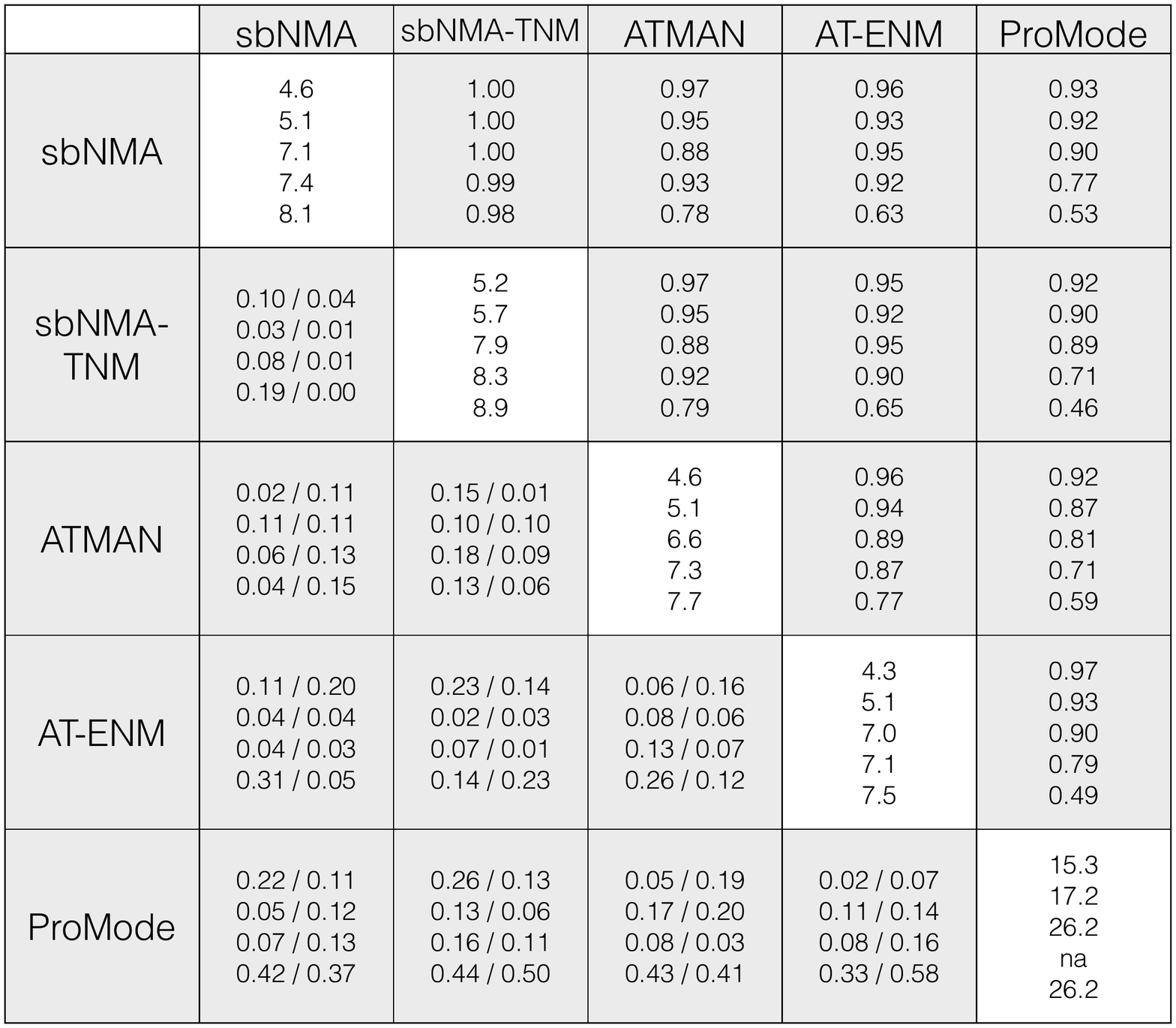}
\caption{Statistics on the cross-comparisons of the slow modes of  1E0W computed using sbNMA, sbNMA-TNM, ATMAN,
ATMAN-ENM, and ProMode. The diagonal blocks list the wave numbers in inverse centimeters of mode $i$, with $i = 1,\dots, 5$.
The upper triangle lists the  atomic correlation of mode $(i,i)$ for $i = 1,\dots, 5$  between each of the  techniques. So, for example,
the correlation of mode 4 between sbNMA-TNM and ATMAN is 0.92.
The lower triangle lists the first off-diagonal correlations for both entries  $\corr{i}{i+1}/\corr{i+1}{i}$, or
more explicitly, for the following pairs of correlations:
$\corr{1}{2}/\corr{2}{1}$;
$\corr{2}{3}/\corr{3}{2}$;
$\corr{3}{4}/\corr{4}{3}$; 
$\corr{4}{5}/\corr{5}{4}$.}
\label{Figure012}
\end{table}
\clearpage

\pagebreak
\begin{figure}[h]
\includegraphics[width=0.8\textwidth]{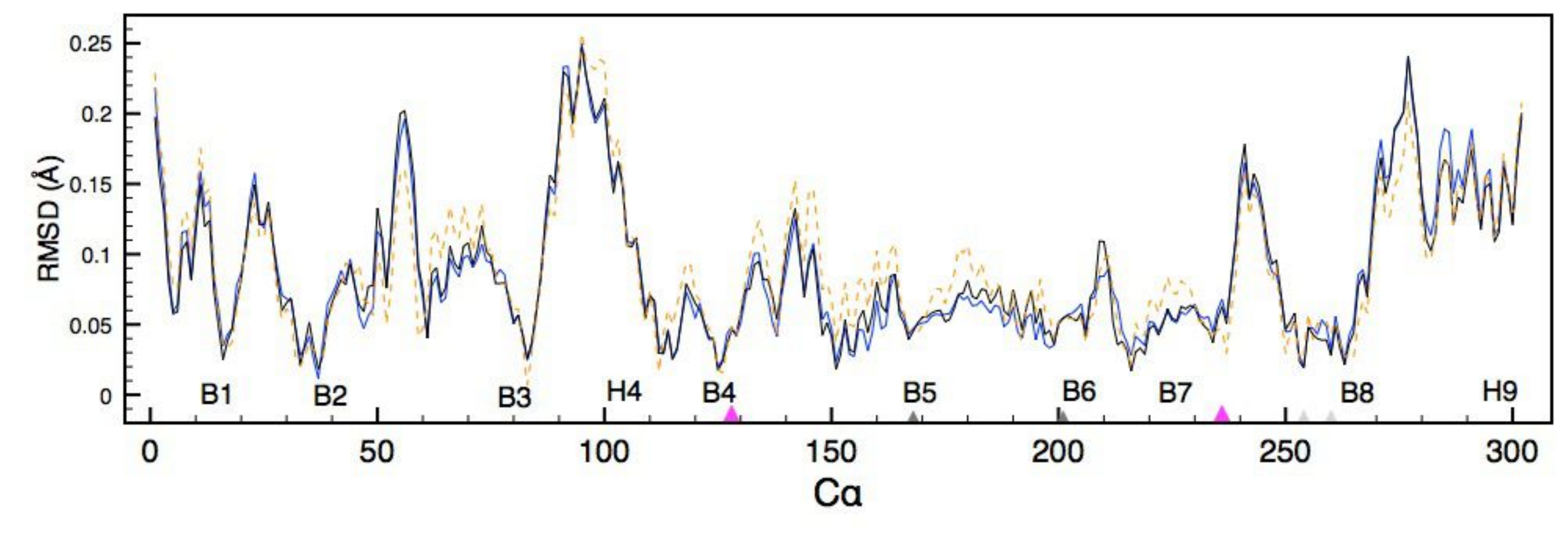}
\label{Figure03}
\end{figure}
\begin{figure}[h]
\includegraphics[width=0.8\textwidth]{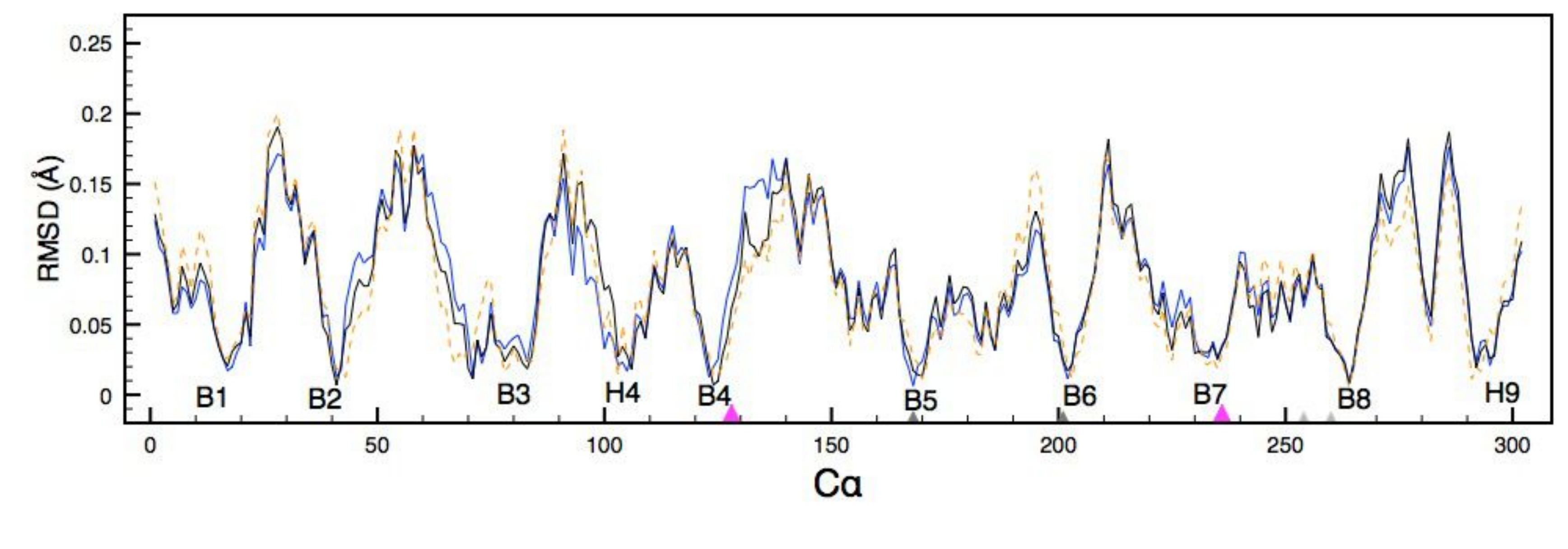}
\label{Figure04}
\end{figure}
\begin{figure}[h]
\includegraphics[width=0.8\textwidth]{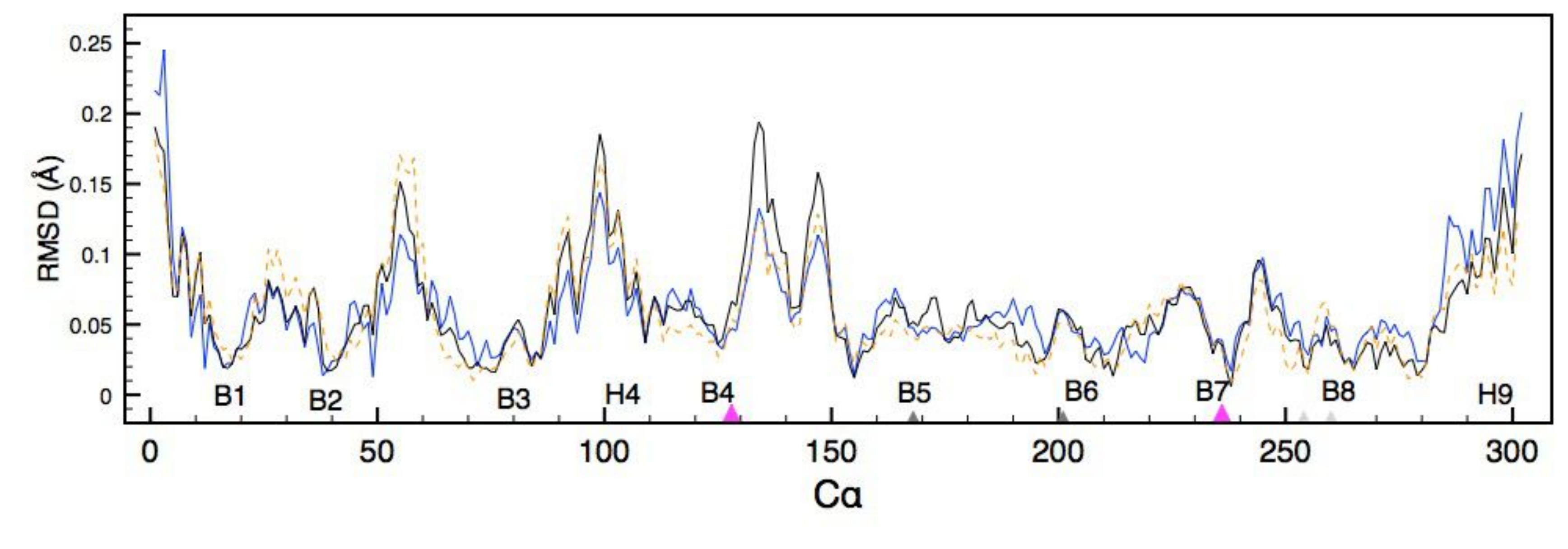}
\caption{The RMSD per $\Ca$ of PDBID 1E0W computed using sbNMA (black), ATMAN (blue) and ProMode (dashed orange) for mode 1 (top) mode 2 (middle) and mode 3 (bottom). The location of the eight $\beta$-strands B1-B8 are labeled as are two of the $\alpha$-helices (H4 and H9). The catalytic residues, Glu128 at the end of B4 and Glu236 at the end of B7 are indicated by magenta pointers. The two pairs of cysteine residues contributing two disulfide bonds are indicated by a pair of dark and light gray pointers.}
\label{Figure05}
\end{figure}
\clearpage

\begin{figure}[h]
\includegraphics[width=0.5\textwidth]{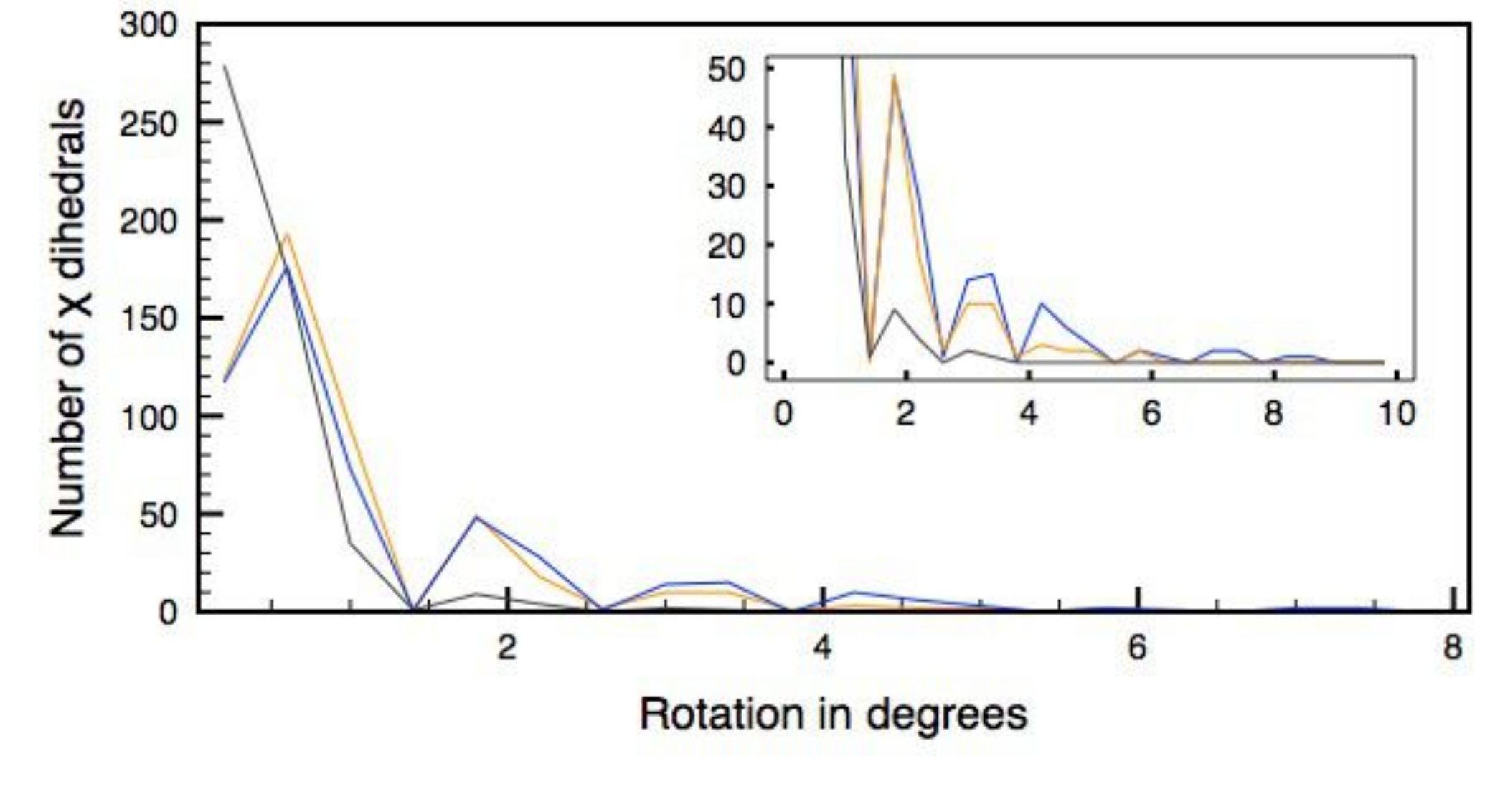}
\caption{A histogram of the rotational flexibility of side chain $\chi$ angles of sbNMA (black), ATMAN (blue) and ProMode (orange). 
For each degree of freedom $\chi_j$  ($j=1,2,\dots,505$) the change $\Delta\chi_j=|\chi_j^1-\chi_j^0|$, between the value of $\chi_j$ for mode~1 (activated to $1\,k_BT$) and its value in the native PDB configuration, is recorded and
the results are collected in 0.4$^\circ$ bins, from 0$^\circ$ to 8$^\circ$  (the $x$-axis). The y-axis reports the number of $\chi$ dihedrals 
in each bin. The inset focuses on the tail end of the distribution, ignoring the central maximum.
}
\label{Figure06}
\end{figure}

\pagebreak
\begin{figure}[h]
\includegraphics[width=0.8\textwidth]{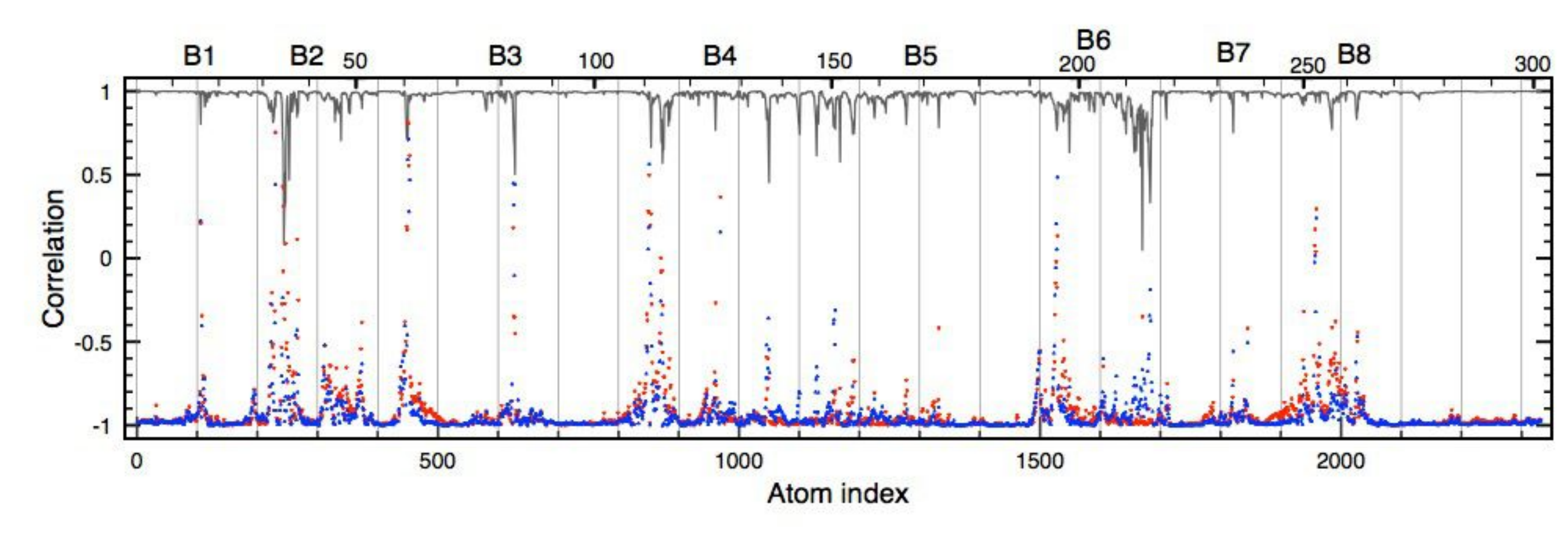}
\label{Figure07}
\end{figure}
\begin{figure}[h]
\includegraphics[width=0.8\textwidth]{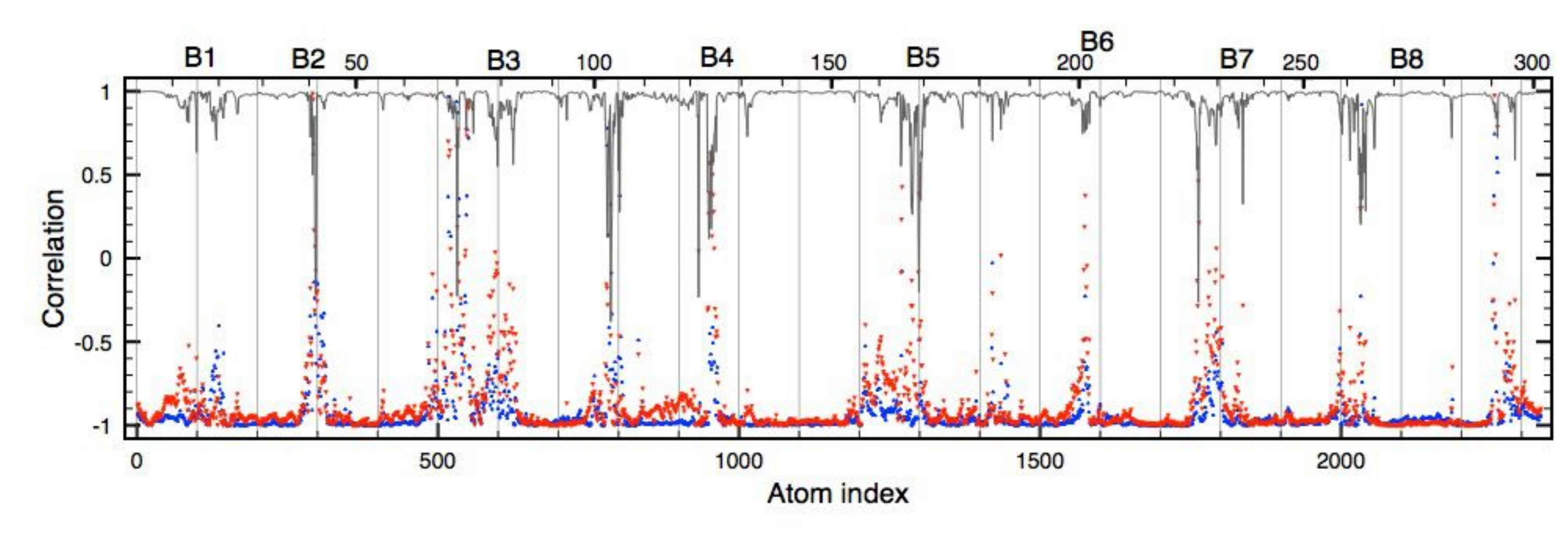}
\label{Figure08}
\end{figure}
\begin{figure}[h]
\includegraphics[width=0.8\textwidth]{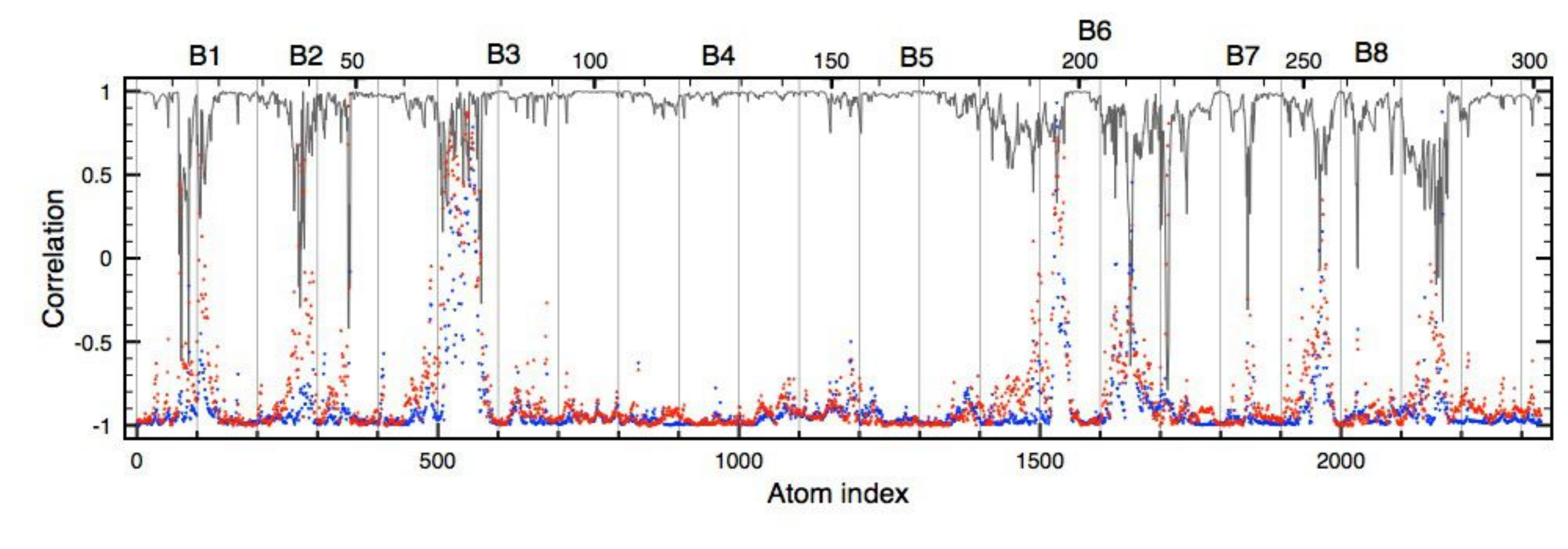}
\caption{The cross-correlations per atom per mode between sbNMA versus ATMAN (black), sbNMA versus ProMode (blue)
and ATMAN versus ProMode (red) for mode 1(top), mode 2 (middle) and mode 3 (bottom). The atom index runs over all heavy
atoms  in the sequence of the PDB coordinate file 1E0W, 
and the residence sequence at the top also indicates the location of the eight $\beta$-strands. For clarity, the black sbNMA-ATMAN data are plotted in-phase (near correlation 1) while the comparisons between sbNMA-ProMode and ProMode-Atman are shown out-of-phase (near correlation -1).
}
\label{Figure09}
\end{figure}
\clearpage

\begin{figure}[h]
\includegraphics[width=0.5\textwidth]{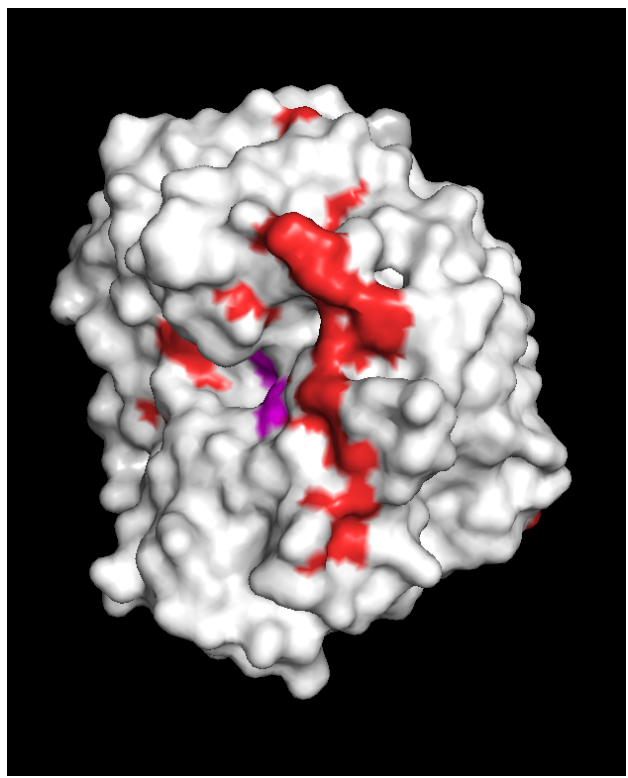}
\caption{Surface representation of 1E0W viewed from the side, with the active site cleft with magenta-colored
 Glu128 and Glu136  facing left and the N and C terminal regions at the top. 
 The coloration reflects the atoms that either
have a very high correlation (white) or a low correlation plus high motility (red) for mode 1 between sbNMA 
and ATMAN.}
\label{Figure10}
\end{figure}
\clearpage

\bibliography{myrefs}

\end{document}